%% file: main.tex
\documentclass[12pt]{article}

\usepackage{amssymb, amsmath}
\usepackage{makeidx}
\usepackage{setspace}
\usepackage{caption}[=v1]
\usepackage{todonotes}
\usepackage{graphicx}
\usepackage[round]{natbib}
\usepackage{dsfont}
\usepackage{color}
\usepackage{topcapt, lscape}
\definecolor{blue}{rgb}{0.4,0,0.4}
\usepackage{epsfig}
\usepackage{float}
\usepackage{url}
\usepackage{lscape}
\usepackage{fullpage}
\usepackage[pagewise]{lineno}
\usepackage[T1]{fontenc}
\usepackage{etoolbox}
\usepackage[toc,header,title, page]{appendix}
\usepackage{titletoc}
\usepackage{charter} % Font: Charter, CM math, CM sans
\usepackage{enumitem}
\usepackage{titlesec}

\titlespacing{\section}{0pt}{\parskip}{-\parskip}
\titlespacing{\subsection}{0pt}{\parskip}{-\parskip}
\titlespacing{\subsubsection}{0pt}{\parskip}{-\parskip}

\usepackage[margin=1.25in]{geometry}
\usepackage{changes}
\usepackage{multirow}
\newcommand{\urlwofont}[1]
{
\urlstyle{same}\url{#1}
}

\definecolor{darkgreen}{rgb}{0.1,0.6, 0.1}

\newcommand*{\TitleFont}{%
      \usefont{\encodingdefault}{\rmdefault}{n}{n}%
      \fontsize{16}{14}%
      \selectfont}

\begin{document}

\title{\TitleFont Learning about Spatial and Temporal \\ Proximity using Tree-Based Methods}

\author{Ines Levin\thanks{Associate Professor, Department of Political Science, University of California at Irvine, 3151 Social Science Plaza, Irvine, CA 92697 (inelevin@uci.edu).}}

\date{February 20, 2022}
\maketitle

\thispagestyle{empty}  % No page number in first page

\begin{abstract} 
\noindent Learning about the relationship between distance to landmarks and events and phenomena of interest is a multi-faceted problem, as it may require taking into account multiple dimensions, including: spatial position of landmarks, timing of events taking place over time, and attributes of occurrences and locations. Here I show that tree-based methods are well suited for the study of these questions as they allow exploring the relationship between proximity metrics and outcomes of interest in a non-parametric and data-driven manner. I illustrate the usefulness of tree-based methods vis-\`a-vis conventional regression methods by examining the association between: (i) distance to border crossings along the US-Mexico border and support for immigration reform, and (ii) distance to mass shootings and support for gun control.
%\\
%Word count: 5,668 \\
\end{abstract}

\vspace{1cm}

\newpage

\pagenumbering{arabic} % Re-start numbering

\setlength{\baselineskip}{21pt}

\setstretch{1.8}

\section{Introduction}

\noindent Factors linked to geography, such as social context within neighborhoods, proximity to events, matters of regional importance, environmental cues, and exposure to local news coverage, may be important drivers of behavior and public opinion. These factors may, additionally, mediate the influence of individual attributes such as party identification, and may thus play a role in explaining important political phenomena, such as the spatial distribution of partisan polarization. Researchers are increasingly taking advantage of the opportunities offered by the advent of Geographic Information Systems (GIS) to incorporate nuanced information about the environment in which individuals find themselves into their data analyses \citep{cho12}. Recent studies have considered influence of spatial proximity to fixed landmarks, such as cities with changing racial makeup \citep{reny18}, mass shootings \citep{newman17}, uprisings of agricultural workers \citep{aidt15}, and national borders \citep{branton07}. Others studies have looked at spatial and temporal distance to series of events taking place over time, such as cycles of protests \citep{branton15, wallace14}. 

Why should researchers care about accurately measuring distance effects? Distance measures may serve as proxy for theoretically-relevant but hard-to-measure factors. For example, distance to state capitals has been used as proxy for familiarity with news about one's own state \citep{carpini94}. But, more crucially, spatial and temporal distances may be substantively important in their own right for understanding certain political phenomena. Theories of diffusion of democratization, international conflict, civil war, and economic prosperity, for instance, highlight the important role of geographical proximity. As noted by \citet[p. 740]{gw2001}, "proximity and distance shape [the] incentives, resources, and constraints" faced by states, and therefore are "critical for understanding how internationalized conﬂicts at the intersection of interstate and civil wars evolve and diffuse." Similar arguments can be made about the theoretical importance of proximity for understanding the diffusion of public opinion and political behavior \citep{enos2017}. Developing a research design for accurately measuring distance effects requires a good understanding of the theoretical relevance of proximity---which will vary between studies depending on the subject matter.

From an applied point of view, learning about the importance of spatial and temporal distances to landmarks and events is a multi-faceted problem, as it may require taking into account multiple dimensions, including: geographical location, timing, and attributes of events in a series of happenings (e.g. number of vehicles going through border crossings, number of individuals participating in protests or attending campaign rallies, number of victims in mass shooting or terrorist attacks, etc.).  Most often, scholars evaluate the relevance of proximity using regression models that assume the existence of a linear, or otherwise known, relationship between selected distance metrics (e.g. distance to the nearest landmark or largest event) and outcomes. As the number of potentially-relevant distance measures grows larger, regression models relying on restrictive parametric assumptions become less adequate, as they may produce inaccurate estimates when the model is overspecified---that is,  when it includes explanatory variables that are not part of the data generating process (DGP)---and may also perform poorly when the model fails to account for complexities in the DGP, such as discontinuities at cutoff points, nonlinearities, and interactions between distance measures.

Researchers often start out by making assumptions about what constitutes substantively meaningful spatial and temporal proximity, or about the relevant size and characteristics of landmarks and events. \citet{aidt15} consider whether a riot occurred within a 10 km radius of each UK Parliamentary constituency, and then repeat the analysis using radiuses of different sizes as robustness check. \citet{wallace14} consider whether a protest occurred within 100 miles of a respondent's location and whether the respondent was interviewed within 30 days of a protest having taken place, distinguishing between proximity to small ($<10,000$ participants) and large protests ($>10,000$ participants).  \citet{branton15}, using similar survey data as \citealt{wallace14}, make a different assumptions about spatial and temporal proximity: whether a respondent was interviewed after the start of the protest cycle, and whether protests occurred in the metropolitan statistical area or county where the respondent resides. 

After making measurement decisions concerning proximity, researchers then proceed to specify linear or generalized linear regression models to explain a certain phenomena of interest. Sometimes regression models include dummy variables used to code observations as near or not-near landmarks or events (with dummy indicators constructed based on criteria such as those described in the previous paragraph). Other times, regression models include raw distance measures (e.g. distance measured in miles) to nearest landmarks or events. Examples of the second approach are \citet{reny18} who include measures of cities' spatial proximity to the nearest `Black growth city;' and \citet{newman17} who include survey respondents' spatial proximity to the nearest mass shootings over a period of time. The former approach (coding binary indicators of proximity) entails assumptions about when an observation can be said to be in the vicinity of a location; the latter approach (including raw distance measures in linear regression models) relies on functional form assumptions about the relationship between distances and outcomes of interest (i.e. that it is linear or has an otherwise known shape). Unless interaction terms are explicitly included in regression models, neither approach allows for the relationship between a given distance measure and the outcome of interest to vary depending on values taken by other explanatory variables (including proximity to other similar landmarks or events).

With these approaches, results can be sensitive to measurement decisions and the choice of how many (and which) distance measures to include in the model. Can we learn about the relationship between distances and outcomes of interest without making restrictive and potentially unrealistic assumptions about how variables relate to each other? Can we relax functional form assumptions and let the data inform us about the substantive significance of spatial and temporal proximity---tell us what it means to `be near' a landmark, when an event can be said to have happened `recently,' the relevant density of events around observations (e.g. whether influence depends on the number of nearby happenings), and inform us about the relevant type of distance metric (e.g. whether distances to the nearest landmarks or most recent events are the only ones that matter, or whether influence depends on landmark or event size, such as number of participants in a protests or victims in violent events)? Recent technological advances offer promising prospects, as in addition to increased data availability, machine learning algorithms and computational resources are now within reach of most practitioners that allow relaxing restrictive modeling assumptions. Tree-based methods, in particular, allow for greater flexibility in modeling the relationship between distances and political behavior, and can therefore help resolve some of the tension that exists between theoretical and applied conceptualizations of the importance of proximity.

\section{Learning about distances with tree-based methods}\label{methodology}

\noindent In an era of widespread data availability, new computational techniques are becoming available that allow researchers to more efficiently explore large and complex data sets and learn from non-traditional types of data, such as textual, network, and spatial information \citep{grimmer15}. Social scientists are increasingly using machine learning techniques---computational methods for extracting information from data sets---to find hidden patterns in data sets with large numbers of observations and predictors \citep{varian14, wallach14}. One such technique are algorithms based on \emph{decision trees} \citep{hastie01, montgomery18}: sets of decision rules with an underlying logic analogous to that of a flowchart for predicting an outcome of interest based values taken by predictive items or features. Approaches based on decision trees---collectively known as tree-based methods---have two key properties that make them suitable to the task of learning about the joint relationship between multiple multi-valued numerical features (such as spatial and temporal distances) and phenomena of interest (such as public opinion on policy issues). Tree-based algorithms: (i) automatically give prominence to predictors based on their ability to explain the outcome of interest; and (ii) do not rely on restrictive parametric assumptions.

To illustrate these points, let us first consider a simple hypothetical example of a single decision tree. Suppose that we were interested in learning about the relationship between distance to the U.S.-Mexico border and a numerical scale of support for restricting undocumented immigration, and that the real relationship between the two variables was such that: (a) Republicans and Independents are more supportive of efforts to curtail undocumented immigration than Democrats; (b) people who live less than 200 km from the border are generally more supportive of similar efforts, specially those living in close proximity to major ports of entry to the country; and (c) party identification is more strongly related to support for immigration reform than distance to the border and to ports of entry. This hypothetical story would be consistent with the tree structure depicted in Figure \ref{fig:figA1} in the Online Appendix. Estimating a decision tree would enable researchers to uncover the distinct relationship between each predictor and the numerical outcome, including: identifying cutoffs in explanatory variables that determine changes in opinion, evaluating the sign of the relationship (e.g. whether values above or below the cutoff generally lead to more or less support), and learning about nonlinearities and interactions between explanatory variables. The tree structure show in Figure \ref{fig:figA1} is useful for illustrative purposes as it has an intuitive interpretation, but is highly simplistic; real-world data can display much more intricate patterns.

Tree-based algorithms are supervised learning procedures. The observed relationship between a set of predictive features (e.g. a collection of distance measures and attributes of nearest/most-recent/largest landmarks or events) and an outcome in training data is used to guide predictions about values of the outcome in unseen (test) data based only on values taken by predictors. A single-tree model, for example, is estimated by recursively partitioning the training data into subsets at each decision node based on whether values of predictors stand below or above a certain threshold (e.g. spatial distance to a given location above or below 200 km), starting with the predictor exhibiting the strongest correlation with the outcome. (Figure \ref{fig:figA2} in the Online Appendix illustrates the data partitioning process for the hypothetical example on support for immigration and distance to the border represented in Figure \ref{fig:figA1}.) The goal of this procedure is to generate decision rules leading to sub-partitions where the dispersion of the outcome variable is relatively low. When further splitting the data into additional sub-partitions risks having an insufficient number of observations in terminal nodes for accurate estimates, or when additional partitioning does not significantly increase homogeneity in values of the outcome variable, the process stops. Subsequently, test data (not used in estimating the model) are passed through the tree structure from top to bottom and assigned predicted values of the outcome of interest.\footnote{\setstretch{1.5} The terms `regression tree' and `classification tree' are reserved for numerical and non-numerical (i.e. categorical) outcomes, respectively.} 

 Single-tree models have important limitations. The tree structure (and therefore, predictions about the influence of spatial and temporal proximity) can be highly sensitive to changes in the data, such as adding (or removing) observations and predictors. Also, single-tree models typically yield coarse predictions (resulting from lack of smoothness in the distribution of predicted values) and consequently may have low predictive accuracy compared to other methods. Both of these issues can be addressed by implementing tree ensemble procedures that combine the predictions of multiple small trees. One popular ensemble method is \emph{random forest} \citep{breiman01}, a technique involving bootstrapping (or \emph{bagging}) of decision trees, where random subsets of the data and predictors are used to estimate an ensemble or \emph{forest} of trees, and where predictions are made by majority voting (in the case of categorical outcomes) or by averaging (in the case of numerical outcomes) across trees in the ensemble. Another popular ensemble method is Bayesian Additive Regression Trees (BART), an algorithm that relies on Markov Chain Monte Carlo (MCMC) simulation to estimate a sum-of-trees model \citep{chipman07}. The larger the number of trees in a tree-ensemble algorithm, the greater the representational and predictive accuracy. BART tends to perform well out of sample, as it is usually applied using conservative settings that favor simple models (with mostly additive effects and few interactions) and help prevent over-fitting. Ensemble learners generally perform better than single-tree learners, but the gain in accuracy comes at a cost, as ensemble algorithms are seen as `black box' procedures where the estimated relationship between predictors and outcomes may not have a straightforward interpretation.

A useful property of tree-based methods for learning about the influence of numerous spatial and temporal distances is that these algorithms perform automatic variable selection, a feature shared with other machine learning algorithms such as regularized linear regression (e.g. LASSO, see \citealp{tibshirani96}).\footnote{\setstretch{1.5} 
Regularized regression procedures such as LASSO penalize large coefficient magnitudes and return coefficient estimates that are shrunk toward zero. Variables whose coefficients are shrunk to exactly zero are effectively dropped from the analysis.} If an explanatory variable (such as a measure of spatial or temporal proximity) does not contribute to increasing homogeneity at any decision node, then it is not used in the tree-growing process. A second property that makes tree-based methods ideal for learning about the potentially complex relationship between distances and outcomes of interest, is that estimation procedures are non-parametric---i.e. do not rely on assumptions about the functional relationship between outcomes and predictive features.  Tree-based algorithms can discover unknown relationships in the data, including highly complex ones involving discontinuities at cutoff points and interactions between explanatory variables. 

\vspace{0.5cm} \noindent \emph{Variable importance}

\noindent A limitation of the conventional implementation of tree-based methods, is that it may be difficult to discern the independent effect of each explanatory variable, particularly when the tree structure is complex or when using ensemble approaches. In this paper, I learn about the relative importance of variables included in ensemble procedures using a permutation-based approach \citep{strobl08} consisting of the following steps:\footnote{\setstretch{1.5} The \emph{randomForest} R package \citep{liaw02} contains functions for calculating permutation-based measures of global and local importance on out-of-bag (OOB) data; that is, on data randomly excluded from each bootstrap sample used in growing individual trees in the ensemble. The advantage of evaluating relative importance on OOB data, rather than on a separate validation set, is that it can be performed \emph{on the fly} (i.e. simultaneously with the model estimation). The procedure I describe is more general. It can be used to calculate permutation-based importance measures on the test data following the implementation of any supervised learner (even when the learning process did involve resampling and therefore did not generate OOB data).}

\begin{enumerate}[nosep]
\itemsep0em 
\item Apply the algorithm to the training data.
\item Calculate predicted values of the outcome variable for the test data and determine the associated $MSE$ \big(denoted $MSE_0$\big).
\item Generate $P$ copies for the test data, where $P$ is the total number of predictors, and for each $p$th copy permute the values of predictor $p$.
\item For each $p$th altered copy of the test data, calculate predicted values of the outcome variable and determined the associated MSE \big(denoted $MSE_{(p)}$\big).
\item For each predictor $p$ calculate $importance_p = \Big(\frac{MSE_{(p)}}{MSE_0} - 1\Big) \times 100$; i.e. percentage increase in MSE in the test data when values of predictor $p$ are permuted.
\end{enumerate}

For each $p$ predictor, $MSE_{(p)}$ can be quite sensitive to the realization of permuted values. To deal with this issue, I repeatedly implemented steps (3-4) considering $K$ independent permutations of values of each predictor $p$, setting $K = 3$. For each $k$th permutation and $p$th predictor I determined $MSE^{k}_{p}$, obtained $MSE_{(p)}=\frac{\sum_{k=1}^{K} MSE^{k}_{p}}{K}$, and then calculated the importance of $p$ using the formula given in step (5). Using multiple-permutations (i.e. setting $K > 1$) leads to more stable estimates of importance than the single-permutation approach.\footnote{\setstretch{1.5} Importance measures produced by steps (1-5) are useful for evaluating the relative relevance of each predictor over the entire data set. As a by-product of the same procedure it is possible to generate \emph{local} importance measures capturing the increase in squared error (SE) for each observation in the test data set when values of each predictor $p$ are permuted.}

Just like simple simulation procedures may be used to make substantively meaningful inferences about quantities of interest in the context of non-linear regression models \citep{king00}, permutation-based measures of variable importance in combination with simulation techniques may be used to extract substantively interesting estimates for tree-based methods and other machine learning algorithms. In sections \ref{app1} and \ref{app2} of this paper I show how these techniques can be used to learn about local marginal effects of changes in distance measures (that is, marginal effects applicable to regions of the data) and illustrate how simulation results can be used to construct visually-revealing representations of the relationship between predictive features and the outcome of interest.

\section{When are tree-based methods a better approach?}

\noindent Before proceeding to the analysis of real-world data, I used Monte Carlo simulation to evaluate and compare the performance of five different modeling mechanisms on synthetic data. The procedures I compared are: linear regression, linear regression with L1-regularization (LASSO), a single regression tree, a regression tree ensemble estimated using a random forest algorithm, and a tree ensemble estimated using BART. In the Online Appendix I present results for methods applicable when the same outcome variable (a 5-point ordered scale) is treated as a categorical variable: ordered logit, LASSO for multinomial outcomes, single classification trees, classification tree ensembles estimated using a random forest, and BART sum-of-trees models for multinomial outcomes. I evaluated all procedures for both small and large sample sizes ($500$ and $5,000$ training observations, respectively, and similar numbers of test observations), for both linear and more complex data generating processes. For each total sample size ($N$) I implemented the following 10-step procedure 1,000 times:

\begin{enumerate}[nosep]
\itemsep0em 
\item Draw $N$ respondents from the 2016 Cooperative Congressional Election Study (excluding CCES respondents from Alaska and Hawaii).
\item Create $N$ hypothetical respondents with similar zip code of residence, basic demographics (sex, education, and age), and party identification on a 3-point scale, as respondents selected in step (1).
\item Randomly select 15 distinct zip codes among all respondent zip codes seen in the 2016 CCES study (excluding Alaska and Hawaii). Set them as location for a series of 15 hypothetical events.
\item Draw 15 random numbers from a uniform distribution with values between 0 and 10. Set them as timing (temporal distance to 2016) for the 15 hypothetical events generated in step (3).
\item Draw 15 random numbers from a uniform distribution with values between 5 and 30. Set them as size of the 15 hypothetical events generated in step (3).
\item For each hypothetical respondent from step (2), calculate the spatial distance to the three most recent and the three largest hypothetical events.
\item Simulate values of an outcome variable for linear and complex data generating processes (described below), based on respondent characteristics imputed in step (2) and distances calculated in step (6).
\item Randomly split the synthetic data set into training and test samples.
\item Model the relationship between synthetic outcomes from step (7), spatial distances, and respondent characteristics, in the training sample, using all procedures.\footnote{\setstretch{1.5} For the three sets of simulations, single decision trees were estimated using R package \emph{rpart} \citep{therneau15a,therneau15b}, with complexity parameter (and therefore number of splits) selected via cross-validation. Random forest was implemented using R package \emph{randomForest} \citep{liaw02}. BART was implemented using R package \emph{BART} \citep{mcCulloch18}. LASSO was implemented using R package \emph{glmnet} \citep{friedman10}, with penalty parameter selected via cross-validation. I used the same packages for applying these algorithms to real-world data in the two applications discussed later in the paper.}.
\item Evaluate the predictive accuracy of each procedure in the test data, by calculating the mean squared error (MSE) (i.e. with $MSE = \frac{\sum{(y - \hat{y})^2}}{n}$, where $y$ and $\hat{y}$ are observed and predicted values of the synthetic outcome, respectively).
\end{enumerate}

Figure \ref{fig:figB1} in the Online Appendix depicts the spatial distribution of respondents and events for a single simulation. At each step in the simulation process I considered two different data generating processes (DGPs):
\begin{itemize}[nosep]
\itemsep0em 
\item \textbf{Linear DGP}: Linear relationship between raw distances measures and the outcome variable. \emph{Expectation}: Linear regression approaches including raw distances as explanatory variables should perform relatively better than tree-based methods. Linear regression procedures including summary indicators of proximity (coded based on the wrong assumptions about what constitutes meaningful proximity) should perform relatively worse than tree-based methods.

\item \textbf{Complex DGP}: Includes thresholds along distance measures, interactions between distances and individual attribute, and nonlinear effects of numerical individual attributes. \emph{Expectation}: Tree-based methods should generally perform better than linear regression approaches. Among tree-based methods, ensemble learners should outperform single-tree learners.
\end{itemize}

\begin{figure}[!h]
	\centering
		\caption{Distribution of mean squared errors}
		\includegraphics[width=1\textwidth]{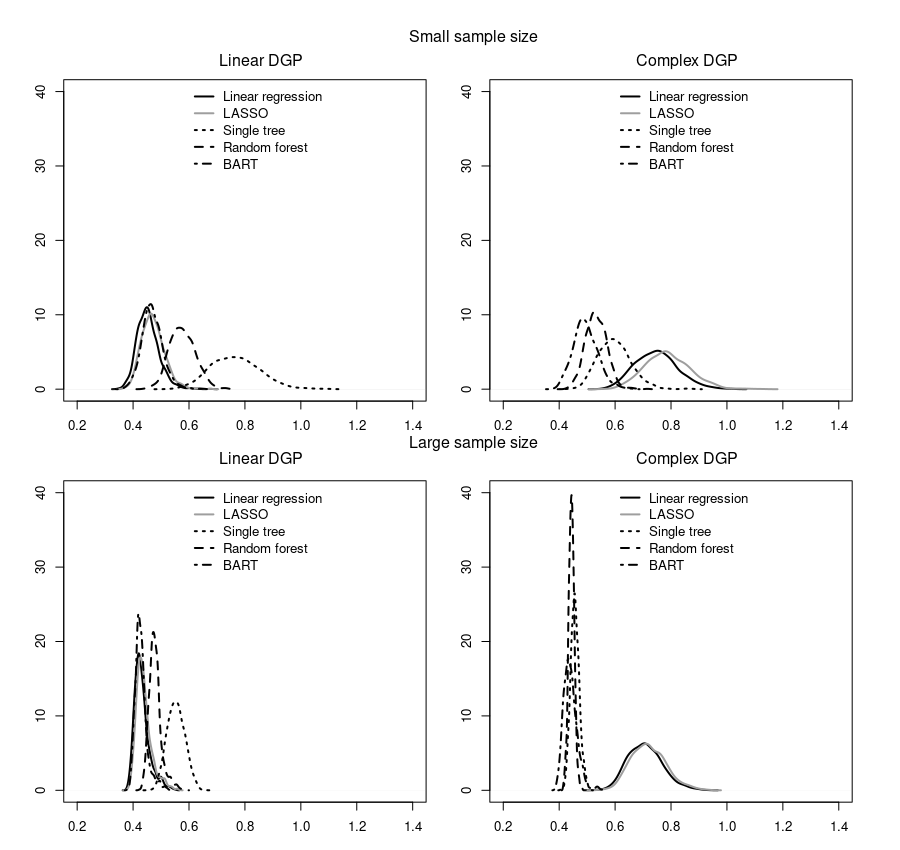}
		\centering{\parbox{5in}{\footnotesize{Note: The figure shows the distribution of MSEs in test data for each procedure (line type), sample size (rows), and type of DGP (columns). The five methods compared in each plot treat the 5-point outcome variable as continuous, and all models include raw distances measured in km.}}}
	\label{fig:fig4}
\end{figure}

The relative performance of each method under linear vs. complex DGPs helps illustrate the advantages of non-parametric tree-based procedures compared to parametric regression approaches. In implementing each estimation procedure, I included all of the variables in either DGP, as well as a number of distance measures and attributes of events that do not form part of either DGP. The inclusion of irrelevant variables helps illustrate the usefulness of automatic variable selection, which here I implement via tree-based methods and LASSO regularization.

Figure \ref{fig:fig4} gives the distribution of MSEs for each estimation approach and DGP.\footnote{\setstretch{1.5} Figures \ref{fig:figB2} and \ref{fig:figB3} in the Online Appendix I show estimated single-tree structures for the case of a linear and complex DGP, respectively.} The two upper plots show results for a small sample size (500 training observations). For a small sample size and linear DGP, the two linear regression approaches (OLS and LASSO) perform about as well as BART, followed closely by random forest, and markedly better than the single-tree learner, which displays much larger MSE over all simulations. For a small sample size and complex DGP, the two tree ensemble approaches perform substantially better than the single-tree learner, and all three tree-based methods outperform the two linear regression procedures. The two lower plots show results for a large sample size (5,000 training observations), where all methods consistently produce more accurate results compared to the small sample size scenario. For a large sample size and linear DGP, linear regression models and BART again outperform other procedures. For a large sample size and complex DGP all tree-based methods, including the single-tree learner, perform considerably better than the linear regression procedures.

When are tree-based methods, then, a better approach? Simulation results confirm that linear regression models including raw distance measures perform relatively well when there is a linear relationship between distances and outcomes of interest. When the regression model does not include raw distance measures but dummy indicators of proximity coded based on the wrong assumptions (e.g. when the dummy indicator assumes that thresholds along distance measures determine substantively-meaningful proximity, but those thresholds do not actually exist in the DGP) linear regression approaches perform poorly even in the context of a linear DGP (see Figure \ref{fig:figB4} in the Online Appendix). Tree ensemble procedures are a generally preferable for explaining outcomes of interest in terms of distance measures, as they produce relatively accurate predictions even in small data sets. Similar conclusions hold for categorical outcomes (see Figures \ref{fig:figB6} and \ref{fig:figB5} in the Online Appendix).

\section{Border crossings and attitudes toward immigration}\label{app1}

\begin{figure}[h!]
	\centering
		\caption{Spatial Distribution of border crossings}
		\includegraphics[width=1\textwidth]{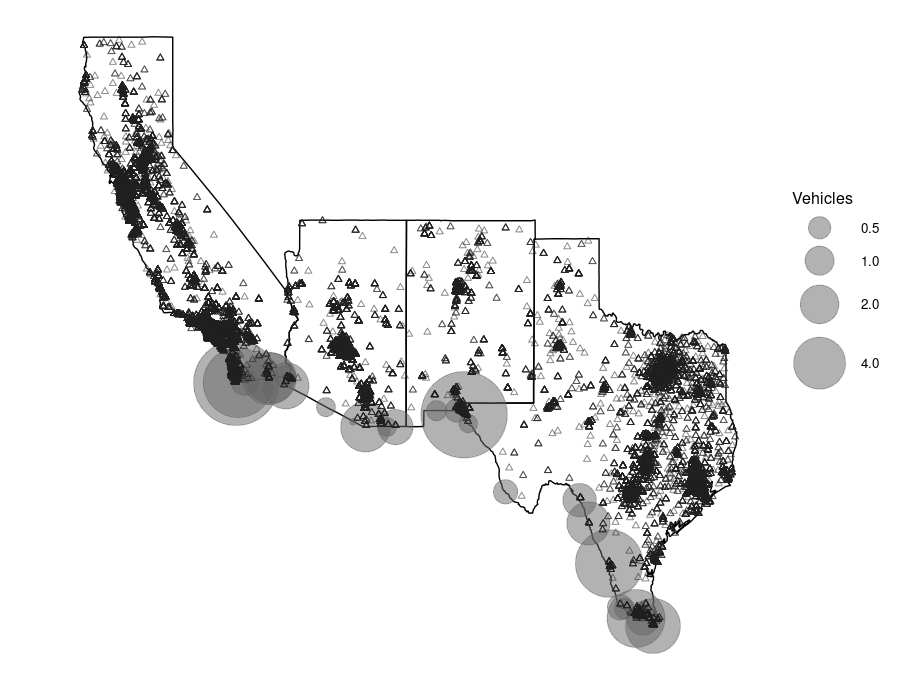}
	\centering{\parbox{5in}{\footnotesize{Note: Triangles represent survey respondents, circles represent border crossings. Larger circles indicate larger border crossings, as measured by the number of vehicles going through the entry point per year.}}}
	\label{fig:sdcrossings}
\end{figure}

\noindent A previous study \citep{branton07} investigated whether Californians residing in closer proximity to the border are more likely to support anti-immigrant ballot initiatives.  Authors found that, indeed, individuals residing in close proximity to the border are more likely to support anti-immigrant stances. In this section I study a related question using data from the 2010, 2012, 2014, and 2016 Cooperative Congressional Election Studies, focusing on 45,700 respondents residing in states located along the US-Mexico border (i.e. Arizona, California, New Mexico, and Texas). I use the methods described in the previous section to explore the relationship between a 5-point scale measuring pro-immigration attitudes and distance to border crossings along the U.S.-Mexico border.\footnote{\setstretch{1.5} The pro-immigration attitudes scale was constructed as an additive index based on expressed support for the following four policies: (1) Grant legal status to all illegal immigrants who have held jobs and paid taxes for at least 3 years, and not been convicted of any felony crimes; (2) Increase the number of border patrols on the U.S.- Mexican border; (3) Allow police to question anyone they think may be in the country illegally; and (4) Fine US businesses that hire illegal immigrants. Larger values along the resulting 5-point scale represent more pro-immigrant stances. Missing values were imputed via using the R package \emph{mice}.} The spatial distribution of border crossings is shown in Figure \ref{fig:sdcrossings}.\footnote{\setstretch{1.5} Data on location and characteristics of border crossings along the US-Mexico border were downloaded from the U.S. Bureau of Transportation Statistics  (BTS) website.} I estimated all models on a subsample of 40,000 randomly selected survey respondents and evaluated the predictive accuracy of each procedure among the remaining 5,700 observations.\footnote{\setstretch{1.5} In addition to measures of spatial distance to the nearest border crossing, I included the number of vehicles (trucks, buses, and personal vehicles, measured in millions) going through the nearest crossing on an annual basis, indicators of party identification (Democrat, Republican, or Independent), gender indicators, age, race indicators (White, Black, Hispanic, Asian, or Other), a 4-point scale of educational attainment, and an indicator of state of residence (Arizona, California, New Mexico, and Texas).}

\vspace{0.5cm} \noindent \emph{Linear regression models}

\noindent One conventional approach to investigating this question would involve estimating a regression model including a measure of distance to the nearest border crossing and a few control variables (e.g. socio-demographic attributes and basic political attitudes of respondents). Model 1 on Table \ref{tab:tabC1} in the Online Appendix shows results for linear regression model along those lines. According to this model, an increase in distance to the border of 100 km is associated with $(0.003 v - 0.016)$ change in support in immigration reform, where $v$ is the size of the nearest border crossing measured in million of vehicles per year. (Since distance was included in thousands of km, this expression is obtained by multiplying the two relevant coefficients by 0.10.) These estimates suggest that the sign of the relationship depends on the size of the nearest crossing; a number ranging between 0.02 and 12.38 million over all border crossings. A reasonable next step for analysts using this approach would involve carrying out simulations to show how the relationship (e.g. point estimates and standard errors associated with marginal effect of distance) vary as a function of size of nearest entry point. An important drawback of this procedure is that the assumption of a linear relationship between the raw distance measure and the outcome of interest might not be realistic. Results might change markedly when alternative specifications (e.g. including polynomial terms) are used.

Another conventional approach to the same question would involve estimating a similar linear regression model with a different formulation of the explanatory variable of interest. Instead of raw distance measures, the model could include a binary indicator proximity to the nearest border crossing constructed based on some discretionary threshold (e.g. a respondents coded as living `near' a border crossing if proximity is lower than 100 km, and as `not near' otherwise). To account for the size of the nearest crossing, analysis could include two separate dummy variables indicating proximity to small and large crossing, respectively, also based on some discretionary criterion. Model 2 in Table \ref{tab:tabC1}, for example, includes binary indicators of proximity to small and large crossings, with proximity determined based on a 250 km threshold and crossing size determined based on a 3 million vehicles/year threshold (results should thus be interpreted relative to people living 250 or more km away from the nearest crossing). These results are qualitatively similar to those produced by Model 1: the larger the size of the nearest entry point, the lower the support for pro-immigration policies. Caution should be exercised in interpreting these findings, however, as results change markedly for different proximity thresholds; the previously apparent relationship is no longer evident in Model 3 in Table \ref{tab:tabC1}, which uses a 100 km threshold.

\vspace{0.5cm} \noindent \emph{Tree-based methods}

\noindent I next used tree-based methods to study the correlates of pro-immigrant attitudes. First, I estimated a single-tree model with a maximum of 10 splits, shown in Figure \ref{fig:figC1} in the Online Appendix. No distance measure or attribute of nearest border crossing was selected by the algorithm in growing this compact tree structure. The tree representation suggests that support for pro-immigration policies is largest among Hispanic Democrats (average support of 3.1); and lowest among Republicans older than 42 who are white or report a race other than black, Hispanic, or Asian, and who were interviewed in 2012 or 2014 (average support of 0.8). I then estimated another single-tree structure with complexity parameter selected as to minimize cross-validation error and found that distance measures and number of vehicles going through the nearest crossings made small differences for predicted attitudes for some combinations of respondent characteristics.

Subsequently, I implemented tree-ensemble algorithms that allow uncovering more nuanced relationships between outcomes and predictive features than single-tree approaches. I estimated a random forest of 200 trees and a BART sum-of-trees model also with 200 trees. When applied to the test data, these algorithms result in lower MSE (1.32 and 1.29, for random forest and BART, respectively) compared to the single tree with complexity selected via cross-validation (MSE $\approx$ 1.40) and linear regression models (MSE $\approx$ 1.40 for all models in Table \ref{tab:tabC1}). 

\begin{figure}[h!]
	\centering
		\caption{Variable importance for explaining pro-immigration attitudes}
		\includegraphics[width=0.9\textwidth]{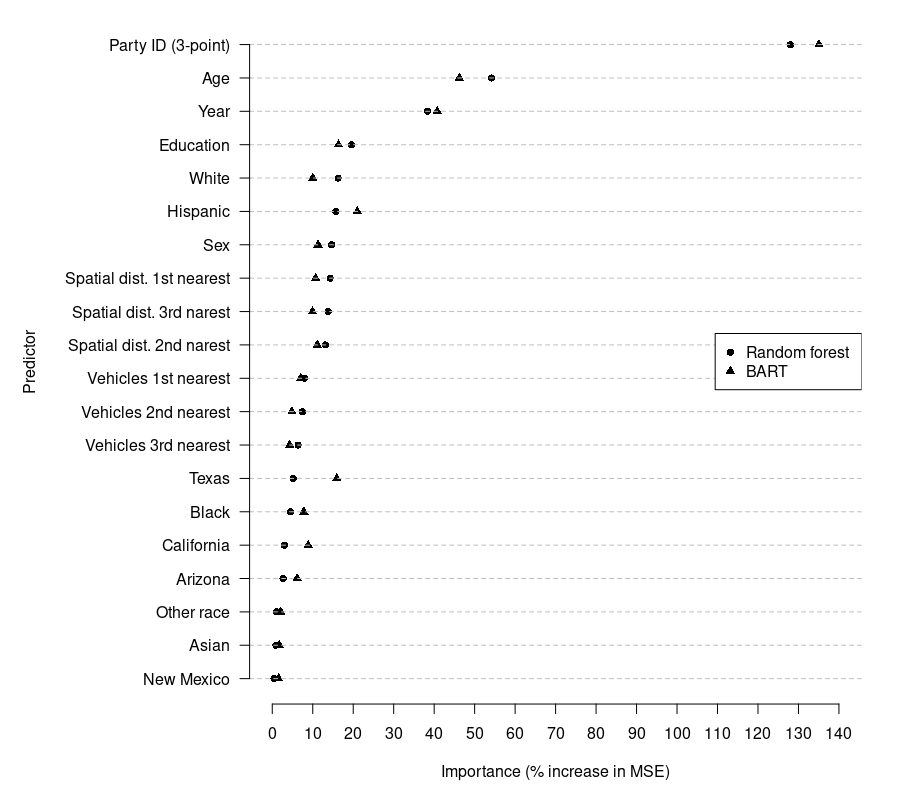}
		\centering{\parbox{5in}{\footnotesize{Note: The plot shows the importance of each predictor, given by the percentage increase in MSE associated with permuting the values of each variable. Predictors are sorted in decreasing order of importance.}}}
	\label{fig:fig10}
\end{figure}

A challenge of tree ensemble algorithms is that the relevance of each predictive feature is not immediately apparent from the standard output produced by most algorithms. Learning about relative variable importance requires post-processing results. Figure \ref{fig:fig10} presents point estimates of relative variable importance calculated using the permutation-based measures described in section \ref{methodology}. These measures suggest that party ID is the most important predictor of pro-immigrant stances (permuting the values of this variable leads to 128\% and 135\% increase in MSE for both random forest and BART, respectively), followed by age (54\% and 46\% increase in MSE, respectively) and interview year (about 38\% and 41\% increase in MSE, respectively). Spatial distances and size of the nearest border crossings have little predictive power; a permutation of values of these variables is associated with less than 15\% increase in MSE for both random forest and BART.

\begin{figure}[h]
	\centering
		\caption{Distance to nearest crossing and pro-immigration attitudes}
		\includegraphics[width=\textwidth]{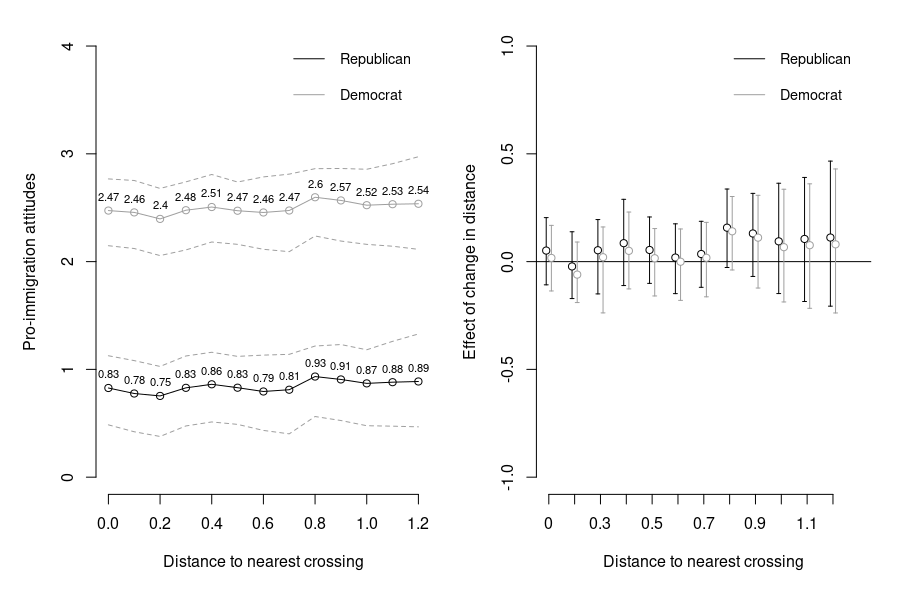}
		\centering{\parbox{5in}{\footnotesize{Note: The plot on the left shows predicted support for pro-immigration policies (vertical axis) for Democratic and Republican hypothetical respondents located at varying distance from the nearest border crossing (horizontal axis, measured in 1,000 km). The plot on the right shows point estimates and 95\% credible intervals for the effect of a change in distance from 100 km (baseline) to the value indicated in the horizontal axis. Results shown in this image correspond to the BART estimation.}}}
	\label{fig:fig6}
\end{figure}

Since tree-based methods allow uncovering deep interactions between predictive features, it may be that the increase in error resulting from permuting the value of any particular predictor is greater for observations displaying certain characteristics. To assess whether this is the case for spatial distances and size of nearest crossings, I also constructed individual-level measures of relative importance for every predictor (i.e. local importance). These measures, on their own, are of limited use, as they tell us nothing about the substantive importance of predictors. To address this I used the following simulation-based procedure to extract quantities of interest: (1) simulated predicted support for pro-immigrant policies for a hypothetical individual for whom distances were determined to have high local importance; (2) simulated predicted support for a Republican with otherwise similar characteristics; and (3) for each of these hypothetical individuals calculated differences in support at each level of distance (0.1 thousand km increases on a 0-1 scale) relative to a baseline distance of 0.1 (i.e. 100 km). The results of this analysis (shown in Figure \ref{fig:fig6}) suggest that even in a region of the data where spatial distance to the nearest border crossing has relatively high importance compared to other predictors, there is no systematic evidence of a relationship between this variable and support for pro-immigration policies, a finding that contrasts markedly with results found with more conventional analyses of the same data.

\section{Mass shootings and support for gun control}\label{app2}

\begin{figure}[h]
	\centering
		\caption{Geographic distribution of mass shootings}
		\includegraphics[width=1\textwidth]{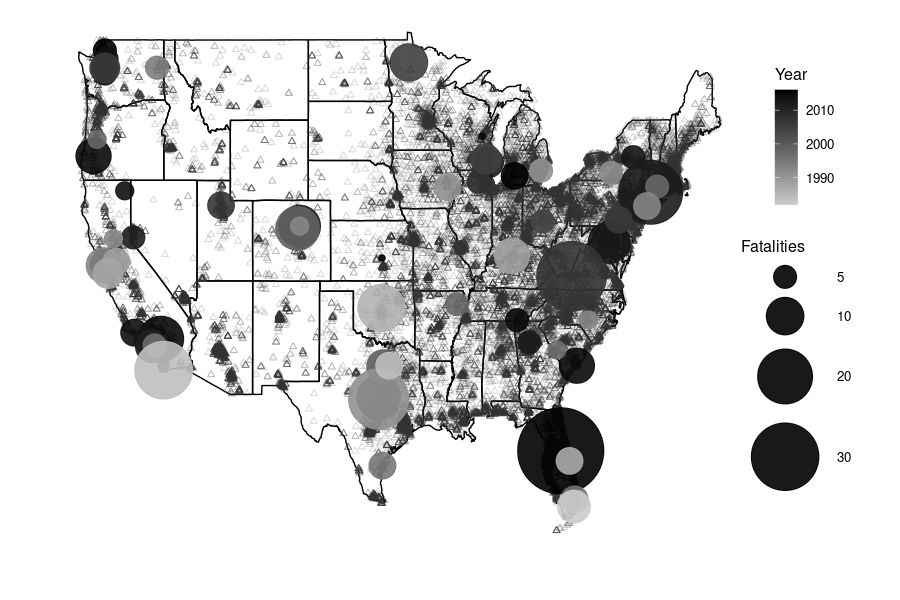}
		\centering{\parbox{5in}{\footnotesize{Note: Triangles represent survey respondents and circles represent mass shootings. Larger circles indicate shootings with a larger number of fatalities. Darker circles indicate more recent events}}}
	\label{fig:fig5}
\end{figure}

\noindent In this section I use data from the 2016 CCES common content and a database of mass shootings in America compiled by \emph{Mother Jones} \citep{follman12} to assess the influence of mass shootings on attitudes toward gun control.\footnote{\setstretch{1.5} Mass shootings are defined as ``attacks that took place in public, in which the shooter and the victims generally were unrelated and unknown to each other, and in which the shooter murdered four or more people'' \citep{cohen14}.} The mass shootings database contains information about the location and magnitude of 83 incidents occurring between 1982 and 2016 (see Figure \ref{fig:fig5}) that I used to calculate a series of spatial and temporal distances to survey respondents. I used a random subsample of 55,000 survey respondents for training purposes and set aside the remaining 9,285 observations to evaluate the predictive accuracy of each procedure.

\vspace{0.5cm} \noindent \emph{Linear regression models}

\noindent I first model support for gun control making similar assumptions about the relevant proximity indicators as \citet{newman17} in their analysis of a similar question using 2010 CCES data: proximity to the nearest shooting, average proximity to the two nearest shootings, and average proximity to the three nearest shootings. I use these distance measures to explain respondents' position along a five-point gun control scale constructed by adding up binary indicators of support for background checks, publishing the names of gun owners, banning assault rifles, and making it more difficult to get conceal carry permit. Table \ref{tab:tabD2} in the Online Appendix presents estimates for three linear regression models of the outcome of interest where the three alternative proximity indicators are included one at a time.\footnote{\setstretch{1.5} In addition to distance measures, the models include party identification indicators, state partisan leaning, a 6-point education scale, age, gender, race indicators, and an indicator of having been the victim of a crime in the year prior to the survey.} Model 1 suggests that a 100 km increase in distance from the nearest mass shooting is associated with 0.05 (s.e. = 0.005) decrease in support for gun control, a negative relationship. (Since distance was included in thousands of km, this number is obtained by multiplying the relevant coefficient by 0.10.) Qualitatively similar results are found for the two alternative proximity indicators (Models 2 and 3 in Table \ref{tab:tabD2}) and in extended models (see Table \ref{tab:tabD3} in the Online Appendix) that also include characteristics of nearest shootings (temporal proximity, number of fatalities, and whether they happened in a school setting).

\vspace{0.5cm} \noindent \emph{Tree-based methods}

\noindent I then estimated a single-tree model with a maximum of 10 splits, shown in Figure \ref{fig:figD2} in the Online Appendix, considering an extended set of distance measures and shooting characteristics (spatial distance, temporal distance, fatalities, and school setting indicator for  each of the three nearest shootings; and spatial distance to each of the three most recent shootings). The tree representation suggests that support for gun control is largest among Democrats older than 48 years old with high levels of education; and lowest among male Republicans who are white or report a race other than Black, Hispanic, or Asian, and who live within 3,700 km from the most recent shooting. Spatial distance to the most recent and second nearest shooting moderate support among respondents with certain combinations of demographic characteristics. Close proximity (< 200 km) to the second nearest shooting (and hence also to the nearest one) is associated with slightly greater support among Independents men who are not black, Hispanic, nor Asian. Largely analogous (though more nuanced) results are found when the complexity parameter is selected as to minimize cross-validation error.

\begin{figure}[h!]
	\centering
		\caption{Variable importance for explaining support for gun control}
		\includegraphics[width=0.9\textwidth]{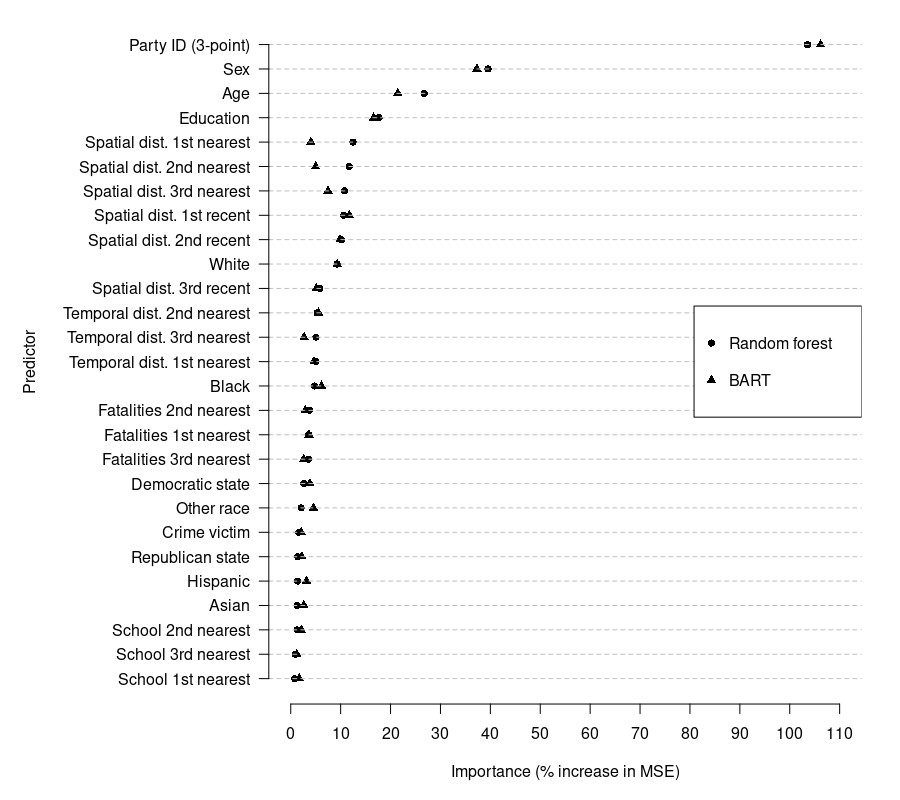}
		\centering{\parbox{5in}{\footnotesize{Note: The plot shows the importance of each predictor, given by the percentage increase in MSE associated with permuting the values of each variable. Predictors are sorted in decreasing order of importance.}}}
	\label{fig:fig7}
\end{figure}

Because of the sensitivity of single-tree models to slight changes in the training data, it is possible that different conclusions (about the association between distance to mass shooting and support for gun control) would have have been drawn for a different random split of the data into training and test sets (or if random subsets of predictive features had been used when considering each splitting decision). To account for this possibility, I then implemented two different ensemble algorithms (random forest and BART with 200 trees each), which also have the advantage of producing smoother (and typically more accurate) predictions. Predictive error is smaller for the ensemble approaches (MSE $\approx$ 1.12 and 1.06 for random forest and BART, respectively) than for the single-tree models (MSE $\approx$ 1.18 for the more complex tree structure), and about equal or smaller than for the linear regressions (MSE $\approx$ 1.11 for all models in Tables \ref{tab:tabD2} and \ref{tab:tabD3}).

Figure \ref{fig:fig7} shows relative importance measures for each predictor, sorted in decreasing order of importance. For both ensemble methods, party identification leads to the greatest increase in MSE (greater than 100\%) upon permutation. Basic demographic attributes including sex, age, and education, are also important predictors of support for gun control (increase in MSE between 17\% and 40\% for the three variables). Spatial distances to the three nearest shootings, and to the three most recent ones, are relatively less important (increase in MSE of 4\% to 12\% for most of these measures), although these measures are found to matter more than other demographics (e.g. race) and attributes of nearest events (such as temporal distance, number of fatalities, and school setting).  
 
\begin{figure}[h]
	\centering
		\caption{Distance to nearest shooting and support for gun control}
		\includegraphics[width=\textwidth]{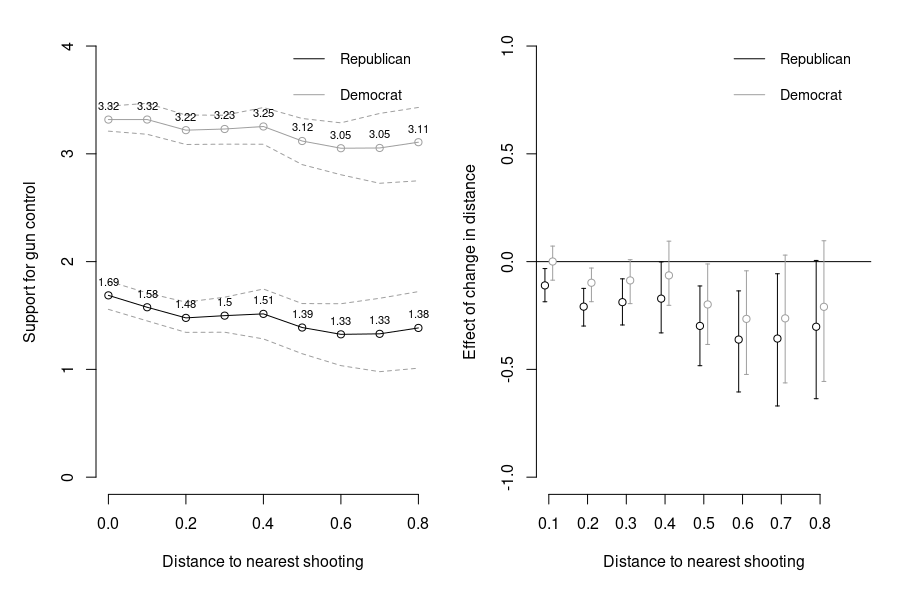}
		\centering{\parbox{5in}{\footnotesize{Note: The plot on the left shows predicted support for gun control (vertical axis) for Democratic and Republican hypothetical respondents located at varying distance from the most recent mass shooting (horizontal axis, measured in 1,000 km). The plot on the right shows point estimates and 95\% credible intervals for the effect of a change in distance from 0 km (baseline) to the value indicated in the horizontal axis. Results shown in this image correspond to the BART estimation.}}}
	\label{fig:guncontrol}
\end{figure}

Lastly, I implemented a procedure similar to the one used in the previous application to: (1) simulate predicted support for gun control for a hypothetical individual for whom spatial distances were determined to have high local importance; (2) simulated predicted support for a Democrat with otherwise similar characteristics; and (3) for each of these hypothetical individuals calculated differences in support at each level of distance (0.1 thousand km increases on a 0-1 scale) relative to a baseline distance of 0.1 (i.e. 100 km). The results of this analysis (shown in Figure \ref{fig:guncontrol}) suggest that individuals located further away from the nearest mass shooting display significantly lower levels of support for gun control. These results are qualitatively similar to those found using conventional regression approaches applied to the same data.
 
\section{Conclusion}

\noindent The paper shows how tree-based methods may be used to explore the relationship between public opinion and distances to landmarks and events. Two important advantages of tree-based methods, compared to conventional regression approaches, are that they: (1) are good at recovering complex relationships in the data, such as discontinuities at cutoff points and interactions between explanatory variables; and (2) perform automatic feature selection, a property that comes useful when researchers are interested in studying the influence of a large number of distance measures. I demonstrated that while conventional regression analyses can produce accurate results when the true relationship between distances and outcomes is fairly simple, they can produce inaccurate and misleading results when the underlying data generating process is complex in unknown ways. 

Not all tree-based methods are equally effective. With single-tree approaches, an in-depth exploration of the relationship between explanatory variables and outcomes of interests requires a large amount of training data. With medium-sized or small data sets, it is often not possible to further grow the tree at specific decision nodes. In those situations, a predictor may not be used for further splitting the data, not out of irrelevance, but due to the small number of training observations reaching terminal nodes. The evidence presented in this paper suggests that researchers would usually be better served using tree ensemble approaches that perform relatively well even in small data sets, such as BART and random forest. While the output of ensemble procedures is more difficult to interpret, I have shown that permutation-based measures of importance in combination with simulation techniques can be used to extract substantively-interesting quantities of interest and alleviate the black box nature of ensemble models.

\newpage
\appendix

\setcounter{page}{1}

\appendixpage

\section*{\Large Supplementary materials for: ``Learning about Spatial and Temporal Proximity using Tree-Based Methods''}

\clearpage

\section*{Supplementary tables and figures}

\setstretch{1.5}

\startlist{lof}{}
\startlist{lot}{}
\vskip-8pt\noindent\hrulefill
\par\noindent\textbf{Figures}
\par\printlist{lof}{l}{}
\par\noindent\textbf{Tables}
\par\printlist{lot}{l}{}
\vskip-8pt\noindent\hrulefill

\clearpage

\input{appendix.tex}

\end{document}

%% file: appendix.tex
\section{Simple hypothetical example}

\setcounter{table}{0}
\renewcommand{\thetable}{A\arabic{table}}

\setcounter{figure}{0}
\renewcommand{\thefigure}{A\arabic{figure}}
 
\begin{figure}[h]
	\begin{centering}
		\caption{Hypothetical example: Distance to the border and support for pro-immigrant policies}
		\includegraphics[width=1\textwidth]{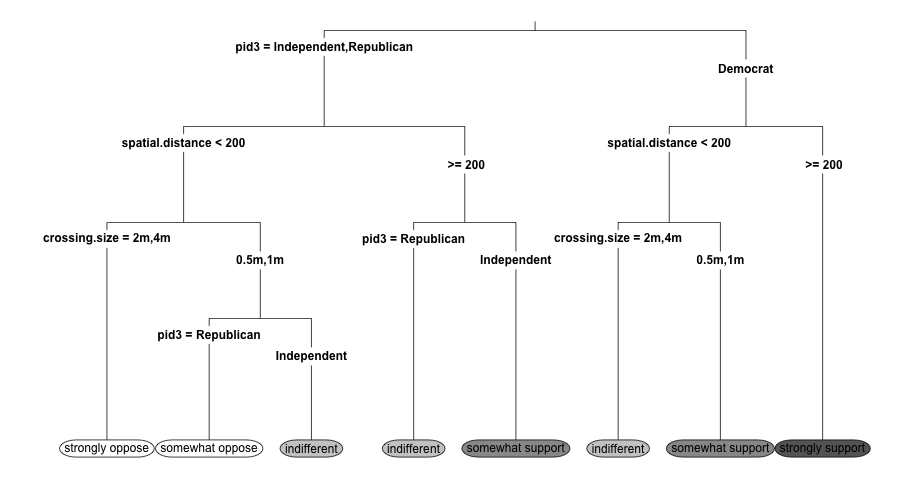}
			\label{fig:figA1}
	\end{centering}
		\centering{\parbox{5in}{\footnotesize{Note: The tree structure indicates how a series of decision rules (i.e. prescriptions for how to split the data based on values of predictive features) should be applied from top to bottom to produce the predictions shown in final nodes (at the bottom of the figure). For instance, this particular tree predicts that support for pro-immigrant policies is largest among Democrats living 200 km or more away from the border, and smallest among Independents and Republicans living less than 200 km from the border and close to a large entry point (with 2-4 million vehicle crossings per year).}}}
\end{figure}

\begin{figure}[h]
	\begin{centering}
		\caption{Hypothetical example: Distance to the border and support for pro-immigrant policies}
		\includegraphics[width=0.8\textwidth]{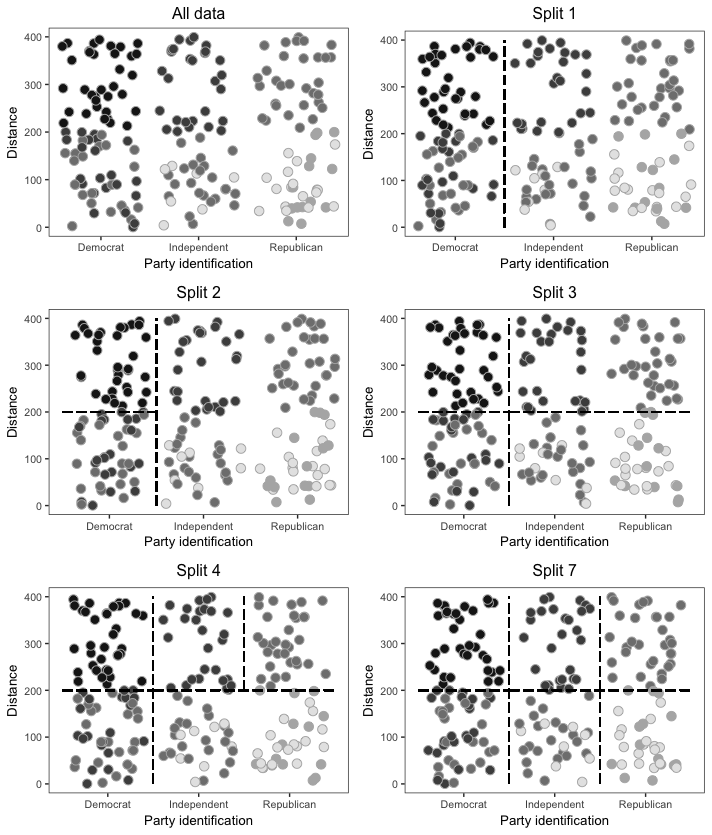}
	\label{fig:figA2}
	\end{centering}
	\centering{\parbox{5in}{\footnotesize{Note: The top-left plot depicts the relationship between party identification (horizontal axis), distance to the nearest border crossing (vertical axis), and support for pro-immigrant policies (circle fill, with darker shades representing greater support), in a hypothetical data set. The rest of the plots depict the data-partitioning process corresponding to the regression tree shown in in Figure \ref{fig:figA1}. First, the data is split based on Party identification (Split 1); among Democrats, and also among Independents and Republicans, it is then split based on distance to the border along a 200 km threshold (Splits 2 and 3, respectively); then among Independents and Republicans living more than 200 km away from the border, it is once again split based on party identification (Split 4). Splits 5-6, not shown, correspond to partitioning done on the basis of size of the nearest border crossing (see Figure \ref{fig:figA1}). Lastly, among Independents and Republicans living less than 200 km away from the border, the data is again split based on party identification (Split 7). Heterogeneity in outcome values within the bottom partitions occurs because of varying size of nearest border crossings (with larger size corresponding to less support).}}}
\end{figure}

\clearpage

\section{Simulation Study}

\setcounter{table}{0}
\renewcommand{\thetable}{B\arabic{table}}

\setcounter{figure}{0}
\renewcommand{\thefigure}{B\arabic{figure}}

\begin{figure}[h]
	\begin{centering}
		\caption{Geographic distribution of simulated chain of events}
		\includegraphics[width=1\textwidth]{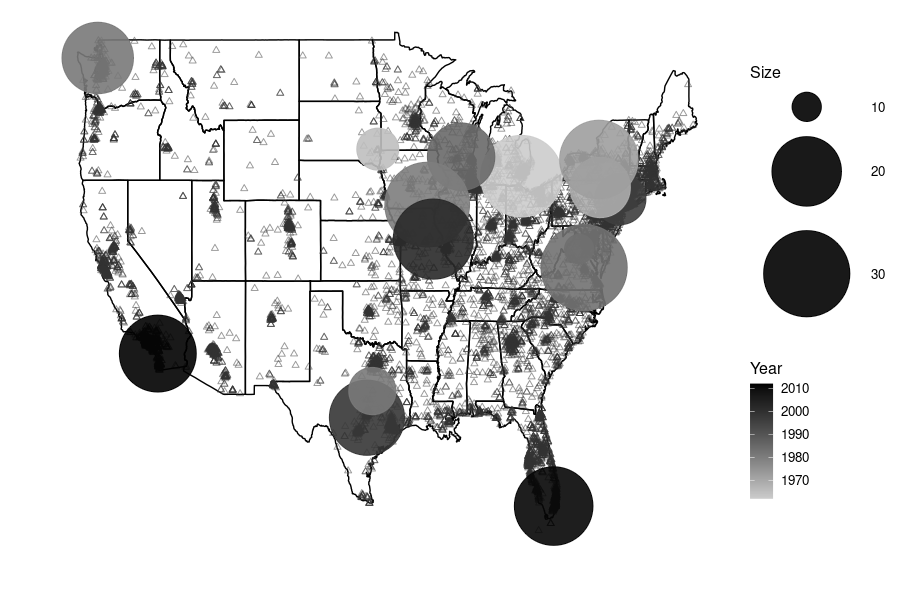}
		\centering{\parbox{5in}{\footnotesize{Note: Triangles represent survey respondents and circles represent hypothetical events. Darker triangles represent greater concentration of survey respondents. Larger circles indicate larger events. Darker circles indicate more recent events.}}}
	\label{fig:figB1}
	\end{centering}
\end{figure}

\clearpage

\begin{figure}[h!]
	\begin{centering}
		\caption{Example of single tree for linear DGP}
		\includegraphics[width=1\textwidth]{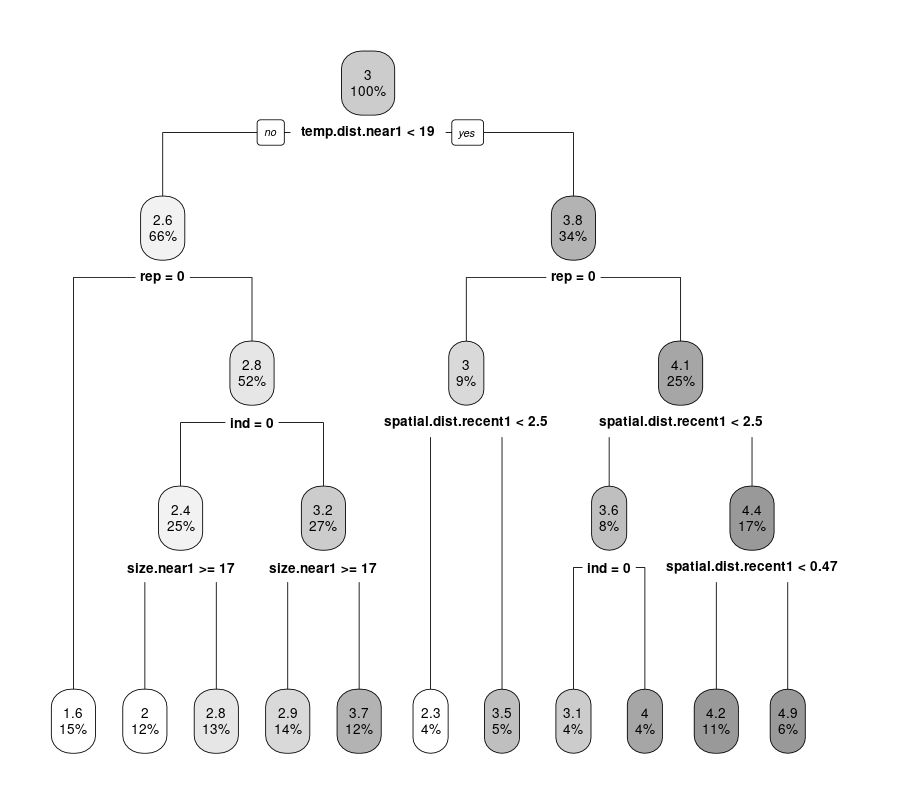}
		\centering{\parbox{5in}{\footnotesize{Note: The figure depicts a single-tree representation corresponding to a single simulation for the linear DGP scenario.}}}
	\label{fig:figB2}
	\end{centering}
\end{figure}

\clearpage

\begin{figure}[ph]
	\centering
		\caption{Example of single tree for complex DGP}
		\includegraphics[width=1\textwidth]{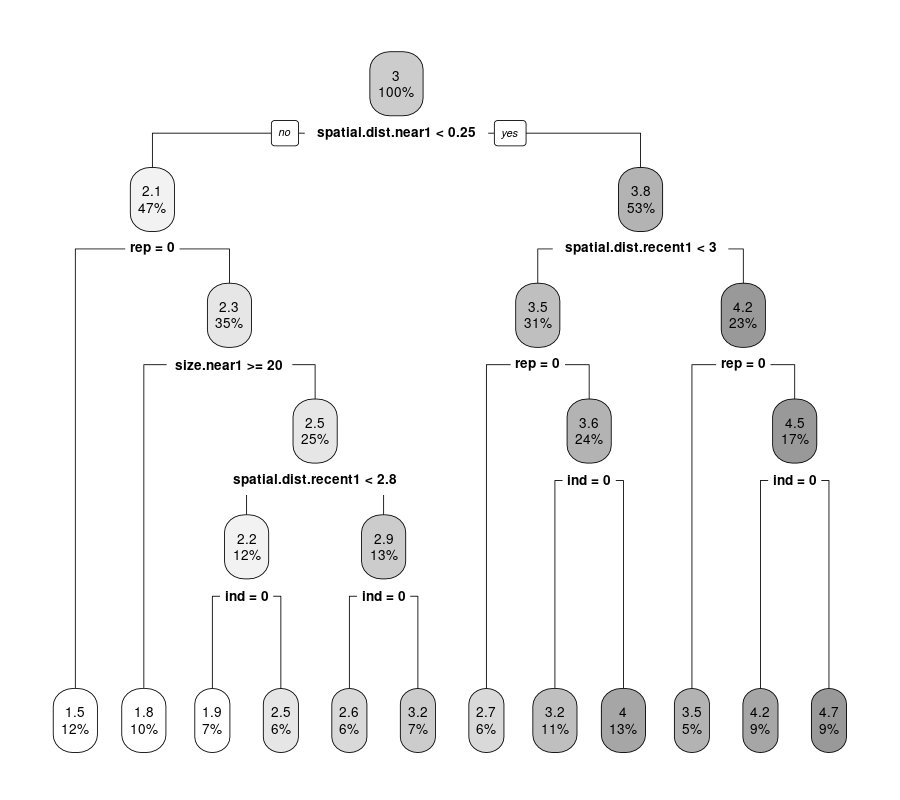}
		\centering{\parbox{5in}{\footnotesize{Note: The figure depicts a single-tree representation corresponding to a single simulation for the complex DGP scenario.}}}
	\label{fig:figB3}
\end{figure}

\begin{figure}[ph]
	\centering
		\caption[Distribution of mean squared errors (binary proximity indicators in linear regression models)]{Distribution of mean squared errors}
		\caption*{(binary proximity indicators in linear regression models)}
		\includegraphics[width=1\textwidth]{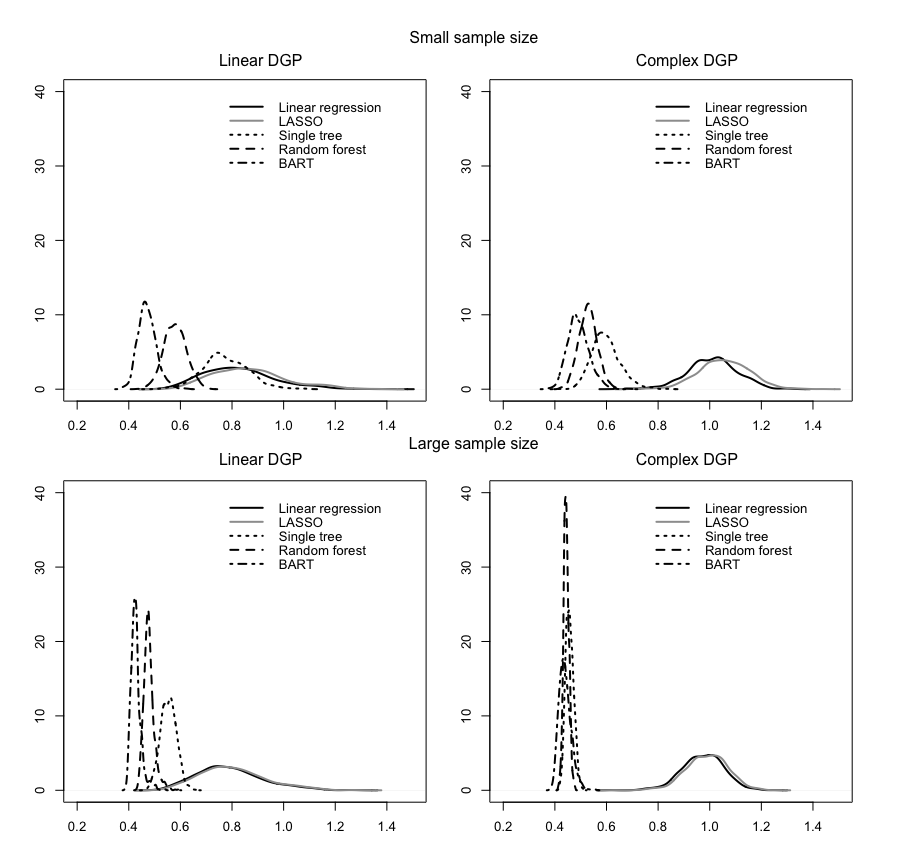}
		\centering{\parbox{5in}{\footnotesize{Note: The figure shows the distribution of MSEs in test data for each procedure (line type), sample size (rows), and type of DGP (columns). The five methods compared in each plot treat the 5-point outcome variable as continuous. In this figure, the linear regression models include binary indicators of proximity constructed under the wrong assumption about the threshold determining close proximity.}}}
	\label{fig:figB4}
\end{figure}

\begin{figure}[ph]
	\centering
		\caption[Distribution of proportion incorrectly classified (categorical outcome variable)]{Distribution of proportion incorrectly classified}
		\caption*{(categorical outcome variable)}
		\includegraphics[width=1\textwidth]{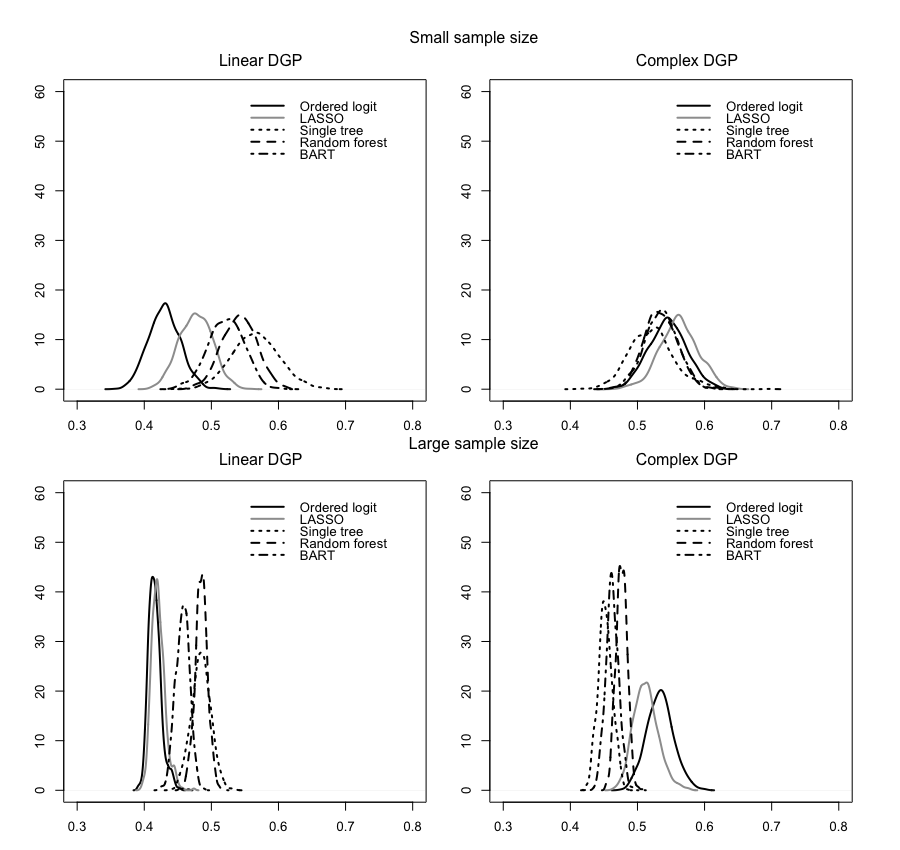}
		\centering{\parbox{5in}{\footnotesize{Note: The figure shows the distribution of proportion of observations incorrectly classified in test data for each procedure (line type), sample size (rows), and type of DGP (columns). The five methods compared in each plot treat the 5-point outcome variable as categorical, and all models include raw distances measured in km.}}}
	\label{fig:figB6}
\end{figure}

\begin{figure}[ph]
	\centering
		\caption[Distribution of mean squared errors (categorical outcome variable)]{Distribution of mean squared errors}
		\caption*{(categorical outcome variable)}
		\includegraphics[width=1\textwidth]{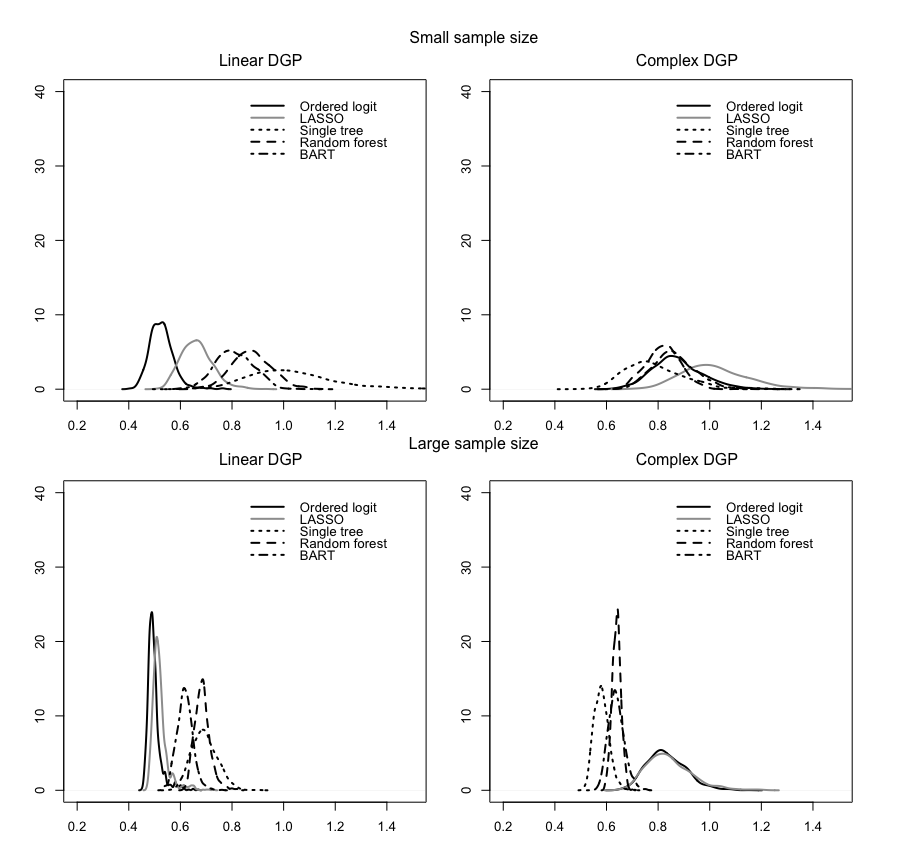}
		\centering{\parbox{5in}{\footnotesize{Note: The figure shows the distribution of MSEs in test data for each procedure (line type), sample size (rows), and type of DGP (columns). The five methods compared in each plot treat the 5-point outcome variable as categorical, and all models include raw distances measured in km. For the purpose of calculating MSEs after applying each model, I treated outcomes as numerical. For a measure of error that does not require making this assumption, see Figure \ref{fig:figB6}.}}}
	\label{fig:figB5}
\end{figure}

\clearpage

\section{Border crossings and support for immigration reform}

\setcounter{table}{0}
\renewcommand{\thetable}{C\arabic{table}}

\setcounter{figure}{0}
\renewcommand{\thefigure}{C\arabic{figure}}

\begin{table}[h!]
\footnotesize
\centering
		\caption{Linear regression models of support for pro-immigration policies}
\begin{tabular}{lcccccc}
 & & & & & &  \\ 
  \hline
 & \multicolumn{2}{c}{Model 1} & \multicolumn{2}{c}{Model 2} & \multicolumn{2}{c}{Model 3} \\ 
 & coef. & s.e. & coef. & s.e. & coef. & s.e. \\ 
  \hline
Intercept & 2.72 & 0.04 & 2.74 & 0.04 & 2.69 & 0.04 \\ 
Independent & -0.81 & 0.01 & -0.82 & 0.01 & -0.82 & 0.01 \\ 
Republican & -1.21 & 0.02 & -1.22 & 0.02 & -1.22 & 0.02 \\ 
Female & 0.18 & 0.01 & 0.18 & 0.01 & 0.18 & 0.01 \\ 
Age & -0.01 & 0.00 & -0.01 & 0.00 & -0.01 & 0.00 \\ 
Education & 0.09 & 0.01 & 0.09 & 0.01 & 0.09 & 0.01 \\ 
Black & 0.08 & 0.02 & 0.08 & 0.02 & 0.07 & 0.02 \\ 
Hispanic & 0.50 & 0.02 & 0.50 & 0.02 & 0.49 & 0.02 \\ 
Asian & 0.08 & 0.03 & 0.08 & 0.03 & 0.08 & 0.03 \\ 
Other & -0.10 & 0.03 & -0.10 & 0.03 & -0.10 & 0.03 \\ 
California & 0.23 & 0.04 & 0.18 & 0.03 & 0.19 & 0.02 \\ 
New Mexico & 0.18 & 0.04 & 0.11 & 0.04 & 0.15 & 0.04 \\ 
Texas & 0.32 & 0.03 & 0.25 & 0.02 & 0.29 & 0.02 \\ 
Year 2012 & 0.03 & 0.02 & 0.03 & 0.02 & 0.03 & 0.02 \\ 
Year 2014 & 0.11 & 0.02 & 0.11 & 0.02 & 0.11 & 0.02 \\ 
Year 2016 & 0.35 & 0.02 & 0.35 & 0.02 & 0.35 & 0.02 \\ 
Dist nearest  & -0.16 & 0.07 &  &  &  &  \\ 
Vehicles nearest & -0.01 & 0.00 &  &  &  &  \\ 
Dist nearest x  Vehicles nearest & 0.03 & 0.01 &  &  &  &  \\ 
Dist nearest $<$ 0.25 \& Vehicles nearest $<=$ 3 &  &  & -0.06 & 0.02 &  &  \\ 
Dist nearest $<$ 0.25 \& Vehicles nearest $>$ 3 &  &  & -0.08 & 0.02 &  &  \\ 
Dist nearest $<$ 0.1 \& Vehicles nearest $<=$ 3 &  &  &  &  & -0.00 & 0.04 \\ 
Dist nearest $<$ 0.1 \& Vehicles nearest $>$ 3 &  &  &  &  & -0.01 & 0.02 \\ 
   \hline				
Adj. $R^2$ & \multicolumn{2}{c}{0.263} & \multicolumn{2}{c}{0.263} & \multicolumn{2}{c}{0.262}  \\ 	
   \hline
\end{tabular}
		\centering{\parbox{5in}{\footnotesize{\vspace{0.5cm} Note: The table gives coefficients and standard errors for three different linear regression models of pro-immigrant attitudes estimated using ordinary least squares. Each model includes a different set of proximity indicators.}}}
	\label{tab:tabC1}
\end{table}

\begin{figure}[p]
	\centering
		\caption{Single-tree model of pro-immigration attitudes (10 splits)}
		\includegraphics[width=\textwidth]{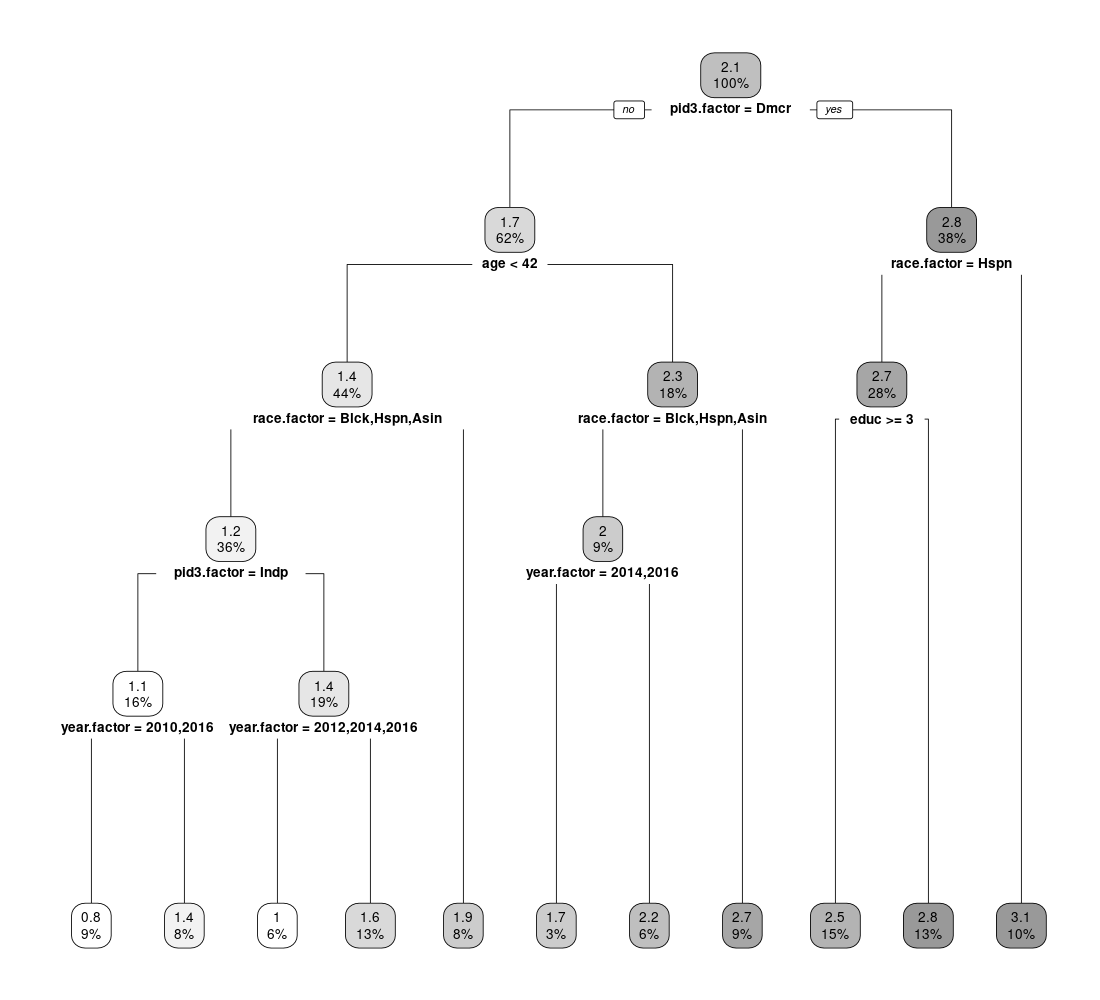}
		\centering{\parbox{5in}{\footnotesize{Note: The figure depicts a single-tree representation with 10 splits for the example given on section \ref{app1} on support for pro-immigration policies.}}}
	\label{fig:figC1}
\end{figure}

\clearpage

\section{Mass shootings and attitudes toward gun control} \label{guncontrol}

\setcounter{table}{0}
\renewcommand{\thetable}{D\arabic{table}}

\setcounter{figure}{0}
\renewcommand{\thefigure}{D\arabic{figure}}

\clearpage

\begin{table}[h]
\footnotesize
\centering
		\caption{Linear regression models of support for gun control}
\begin{tabular}{lcccccc}
 &  & & & &  &  \\ 
  \hline
 & \multicolumn{2}{c}{Model 1} & \multicolumn{2}{c}{Model 2} & \multicolumn{2}{c}{Model 3} \\ 
 & coef. & s.e. & coef. & s.e. & coef. & s.e. \\ 
  \hline
Intercept & 2.587 & 0.022 & 2.617 & 0.023 & 2.628 & 0.023 \\ 
Independent & -0.652 & 0.011 & -0.652 & 0.011 & -0.652 & 0.011 \\ 
Republican & -1.078 & 0.012 & -1.077 & 0.012 & -1.077 & 0.012 \\ 
Age & 0.003 & 0.000 & 0.003 & 0.000 & 0.003 & 0.000 \\ 
Education & 0.067 & 0.005 & 0.067 & 0.005 & 0.068 & 0.005 \\ 
Black & 0.056 & 0.015 & 0.053 & 0.015 & 0.055 & 0.015 \\ 
Hispanic & 0.038 & 0.015 & 0.037 & 0.015 & 0.035 & 0.015 \\ 
Asian & 0.288 & 0.026 & 0.286 & 0.026 & 0.285 & 0.026 \\ 
Other & -0.302 & 0.025 & -0.301 & 0.025 & -0.301 & 0.025 \\ 
Democratic state & 0.063 & 0.011 & 0.049 & 0.011 & 0.045 & 0.011 \\ 
Republican state & -0.051 & 0.012 & -0.048 & 0.012 & -0.053 & 0.012 \\ Crime victim & -0.173 & 0.020 & -0.172 & 0.020 & -0.173 & 0.020 \\ 
Spat. dist. nearest & -0.542 & 0.052 &  &  &  &  \\ 
Avg. spat. dist. 2 nearest &  &  & -0.569 & 0.050 &  &  \\ 
Avg. spat. dist. 3 nearest &  &  &  &  & -0.517 & 0.047 \\ 
   \hline				
Adj. $R^2$ & \multicolumn{2}{c}{0.196} & \multicolumn{2}{c}{0.197} & \multicolumn{2}{c}{0.197}\\ 	
   \hline
\end{tabular}
		\centering{\parbox{5in}{\footnotesize{\vspace{0.5cm} Note: The table gives coefficients and standard errors for three different linear regression models of support for gun control estimated using ordinary least squares. Each model includes a different distance measure.}}}
	\label{tab:tabD2}
\end{table}

\begin{table}[h]
\footnotesize
\centering
		\caption[Linear regression models of support for gun control (controlling for shooting characteristics)]{Linear regression models of support for gun control}
		\caption*{(controlling for shooting characteristics)}
\begin{tabular}{lcccccc}
 &  & & & &  &  \\ 
  \hline
 & \multicolumn{2}{c}{Model 4} & \multicolumn{2}{c}{Model 5} & \multicolumn{2}{c}{Model 6} \\ 
 & coef. & s.e. & coef. & s.e. & coef. & s.e. \\ 
  \hline
  
Intercept & 2.407 & 0.029 & 2.262 & 0.038 & 2.324 & 0.037 \\ 
Independent & -0.652 & 0.011 & -0.651 & 0.011 & -0.651 & 0.011 \\ 
Republican & -1.078 & 0.012 & -1.075 & 0.012 & -1.075 & 0.012 \\ 
Female & 0.446 & 0.009 & 0.447 & 0.009 & 0.447 & 0.009 \\ 
Age & 0.004 & 0.000 & 0.004 & 0.000 & 0.004 & 0.000 \\ 
Education & 0.067 & 0.005 & 0.067 & 0.005 & 0.068 & 0.005 \\ 
Black & 0.043 & 0.015 & 0.050 & 0.015 & 0.058 & 0.015 \\ 
Hispanic & 0.040 & 0.015 & 0.040 & 0.015 & 0.040 & 0.015 \\ 
Asian & 0.299 & 0.026 & 0.296 & 0.026 & 0.292 & 0.026 \\ 
Other & -0.293 & 0.025 & -0.293 & 0.025 & -0.295 & 0.025 \\ 
Democratic state & 0.098 & 0.011 & 0.048 & 0.012 & 0.021 & 0.012 \\ 
Republican state & -0.033 & 0.012 & 0.007 & 0.013 & 0.012 & 0.014 \\ 
Crime victim & -0.166 & 0.020 & -0.164 & 0.020 & -0.166 & 0.020 \\ 
Spat. distance nearest & -0.468 & 0.053 &  &  &  &  \\ 
Temp. distance nearest & 0.000 & 0.001 &  &  &  &  \\ 
pat. dist. most recent & 0.050 & 0.005 &  &  &  &  \\ 
Fatalties nearest & -0.000 & 0.001 &  &  &  &  \\ 
School setting nearest & -0.005 & 0.013 &  &  &  &  \\ 
Avg. spat. dist. 2 nearest &  &  & -0.480 & 0.051 &  &  \\ 
Avg. temp. dist. 2 nearest &  &  & 0.001 & 0.001 &  &  \\ 
Avg. spat. dist. 2 most recent &  &  & 0.141 & 0.013 &  &  \\ 
Avg. fatalities 2 nearest &  &  & 0.000 & 0.001 &  &  \\ 
Prop. school setting 2 nearest &  &  & -0.022 & 0.019 &  &  \\ 
Avg. spat. dist. 3 nearest &  &  &  &  & -0.461 & 0.047 \\ 
Avg. temp. dist. 3 nearest &  &  &  &  & 0.001 & 0.001 \\ 
Avg. spat. dist. 3 most recent &  &  &  &  & 0.137 & 0.014 \\ 
Avg. fatalities 3 nearest &  &  &  &  & -0.000 & 0.001 \\ 
Prop. school setting 3 nearest &  &  &  &  & -0.030 & 0.022 \\ 
   \hline				
Adj. $R^2$ & \multicolumn{2}{c}{0.198} & \multicolumn{2}{c}{0.199} & \multicolumn{2}{c}{0.198}\\ 	
   \hline
\end{tabular}
		\centering{\parbox{5in}{\footnotesize{\vspace{0.5cm} Note: The table gives coefficients and standard errors for three different linear regression models of support for gun control estimated using ordinary least squares. Each model includes a different distance measure and indicators of shooting characteristics.}}}
	\label{tab:tabD3}
\end{table}

\clearpage

\begin{figure}[p]
	\centering
		\caption{Single-tree model of attitudes toward gun control (10 splits)}
		\includegraphics[width=\textwidth]{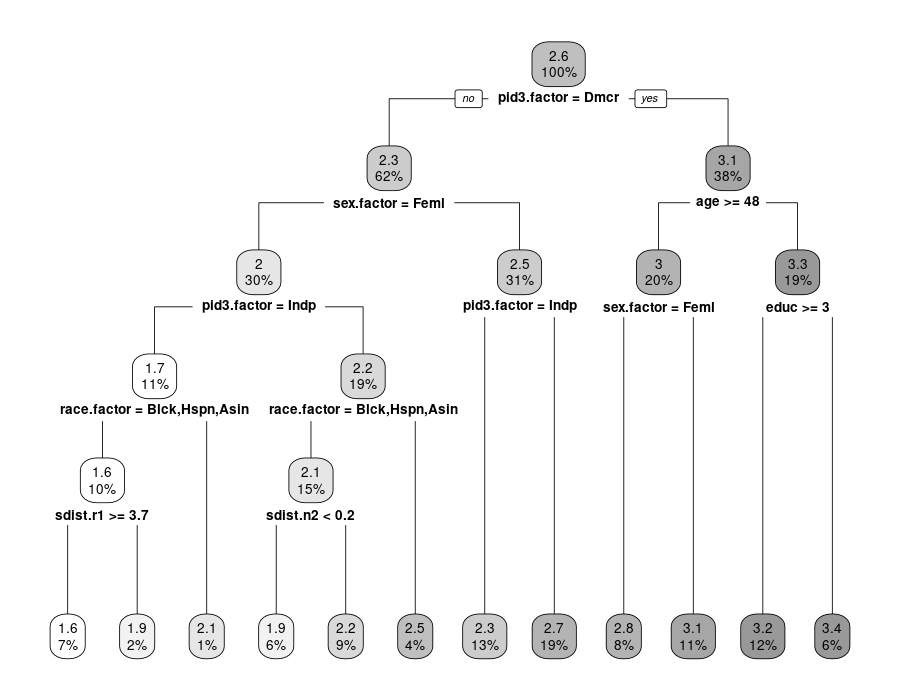}
		\centering{\parbox{5in}{\footnotesize{Note: The figure depicts a single-tree representation with 10 splits for the example given on section \ref{app2} on support for gun control.}}}
	\label{fig:figD2}
\end{figure}